\documentclass[useAMS,usenatbib]{mn2e}
\usepackage{graphicx}
\usepackage{multirow}

\usepackage{hyperref}

\title[Connecting the most massive galaxies at $z < 3$]{Tracing galaxy populations through cosmic time: A critical test of methods for connecting the same galaxies between different redshifts at $z < 3$}

\author[C.J. Mundy, C.J. Conselice and J.R. Ownsworth]{Carl J. Mundy$^{1}$\thanks{E-mail:
\texttt{carl.mundy@nottingham.ac.uk}}, Christopher J. Conselice$^{1}$ and Jamie R. Ownsworth$^{1}$ \\
$^{1}$School of Physics and Astronomy, University of Nottingham, Nottingham, NG7 2RD, UK}

\date{\today}

\pagerange{\pageref{firstpage}--\pageref{lastpage}} \pubyear{2015} 

\begin{document} 

\maketitle

\label{firstpage}
\begin{abstract}
Connecting galaxies with their descendants (or progenitors) at different redshifts can yield strong constraints on galaxy evolution. Observational studies have historically selected samples of galaxies using a physical quantity, such as stellar mass, either above a constant limit or at a constant cumulative number density. Investigation into the efficacy of these selection methods has not been fully explored. Using a set of four semi-analytical models based on the output of the Millennium Simulation, we find that selecting galaxies at a constant number density (in the range $-4.3 < \log\ n\ [\mathrm{Mpc}^{-3}\ h^{3}] < -3.0$) is superior to a constant stellar mass selected sample, although it still has significant limitations. Recovery of the average stellar mass, stellar mass density and average star-formation rate is highly dependent on the choice of number density but can all be recovered to within $<50\%$ at the commonly employed choice of $\log\ n\ [\mathrm{Mpc}^{-3}\ h^{3}] = -4.0$, corresponding to  $\log M_\odot / h \sim 11.2$ at $z=0$, but this increases at lower mass limits. We show that there is a large scatter between the location of a given galaxy in a rank ordering based on stellar mass between different redshifts. We find that the inferred velocity dispersion may be a better tracer of galaxy properties, although further investigation is warranted into simulating this property. Finally, we find that over large redshift ranges selection at a constant number density is more effective in tracing the progenitors of modern galaxies than vice-versa.
\end{abstract}


\section{Introduction}
\noindent In the now commonly accepted paradigm, galaxies form in the gravitational wells of collapsed cold dark matter halos, which themselves are seeded by primordial quantum fluctuations in the Universe's first moments. In this hierarchical picture of galaxy formation and evolution, galaxies build up stellar mass through both in-situ star-formation and galaxy mergers, where more massive galaxies merge with smaller systems which result in more massive descendants. Over time, these processes produce the array of galaxies and the environments within which they are observed. In order to understand how galaxies form and evolve, the evolution in their properties (e.g. stellar mass, size) must be observed. As the most massive, and thus the brightest, galaxies are the easiest systems to observe out to high redshift, it is these galaxies and their properties that observations attempt to study. This has typically been achieved by selecting galaxies in two ways.

Historically, selecting galaxies across a redshift range of interest using a constant stellar mass cut has been used to study the evolution of the most massive galaxies \citep[e.g.][]{Conselice2003,Mortlock2013}. Use of this selection method intrinsically assumes galaxies have more or less been a passively evolving population from high redshift. However, processes such as major galaxy mergers and bursts of star-formation interfere with these assumptions and contaminate the selection by changing the rank order of galaxies. Thus, the wider the redshift range this method is applied to, the less accurately it may trace the galaxies of the original selection.

Selecting galaxies at a constant cumulative comoving number density, when ranked by some physical property such as stellar mass or luminosity, has proven a popular alternative in the recent literature when observing both field and cluster galaxy evolution \citep[e.g.][]{Lin2013}. Using this technique, the averaged star-formation history of a constant number density selected sample of galaxies over the redshift range $3 < z < 8$ has been shown to be able to account for the stellar mass growth of these galaxies \citep{Papovich2010}. The average stellar mass of the most massive galaxies (those with $\log M_{\star} > 11.0$ at $z=0$, and $\log n\ [\mathrm{Mpc}^{-3}] = -4.0$) has been found to increase by a factor of $\sim 4$ over the redshift range $0.3 < z < 3$. However, the integrated star-formation history appears unable to account for the growth in stellar mass at $z < 1.5$. Therefore, the influence of both major and minor galaxy mergers is required to account for this discrepancy at low redshift \citep{Ownsworth2014}.

Furthermore, studies have observed the evolution in H$\alpha$ equivalent width, structural properties and colours of galaxy populations selected at various number densities \citep{Fumagalli2012,Marchesini2014}. Stellar mass measured inside a radius of $r = 5$kpc on stacked images of massive galaxies (selected at $\log n\ [\mathrm{Mpc}^{-3}] = -3.7$ corresponding to galaxies with $\log\ M^*\ [M_\odot] > 11.4$ and $\log\ M^*\ [M_\odot] > 11.1$ at $z=0.1$ and $z=2.0$ respectively) is found to be approximately constant over the redshift range $0.6 < z < 2.0$. On the other hand, the stellar mass content beyond this radius is found to increase by a factor of $\sim 4$ \citep{vanDokkum2010}. \citet{Conselice2013} compare the derived gas fractions of massive galaxies ($M_* > 10^{11}\ \mathrm{M}_\odot$) with their star-formation histories in the redshift range $1.5 < z < 3$, selecting galaxies at a merger-adjusted constant number density. They conclude that gas accretion is the dominant source of observed stellar mass production for these galaxies over this redshift range.

Investigation into the efficacy of either selection method has not been fully explored. Numerical calculations presented in \citet{vanDokkum2010} suggest the influence of galaxy mergers has little influence on stellar mass growth when measured using a constant number density selected sample. \citet{Papovich2010} used dark matter halo merger trees from the Millennium Simulation\footnote{\url{http://www.mpa-garching.mpg.de/galform/virgo/millennium/}} \citep[MS;][]{Springel2005,Lemson2006} to show that the recovery fraction of descendant halos at redshifts $3 < z < 8$ is $\sim 50\%$. \citet{Behroozi2013} found a small change in the cumulative number density of the most massive ($M_* > 10^{11.5} \mathrm{M}_\odot$) $z=0$ progenitor galaxies of $+0.22$ dex per unit $\Delta z$. Furthermore, they find that this change and thus the mass histories of descendants and progenitors are different. More directly, \citet{Leja2013} used the \citet[][G11]{Guo2011} semi-analytical model (SAM) applied to the MS in order to test the validity of the underlying assumptions of constant number density selection. They find that, within this particular SAM, the median stellar mass of descendant galaxies can be recovered over the redshift range $0 < z <3$ to within $\sim 40\%$ of the true value. Corrections for stellar mass growth rate scatter, galaxy mergers and quenching are found to reduce this discrepancy to within $\sim 12\%$ - well within typical observational error attributed to the calculation of stellar masses. These results, however, are model and cosmology dependent and are sensitive to the dark matter merger trees and the recipes used to determine galaxy properties. How the stellar mass is calculated is a prime example. Different methods of calculating this may introduce different levels of scatter into the rank order of galaxies across redshifts. Sensibly investigating the efficacy of these techniques requires a mixture of SAMs, merger trees and cosmology to gauge the amount of variance in the results.

While stellar mass is used in the studies mentioned previously, it may not be the most appropriate property with which to rank and select galaxies in order to trace their properties. Increasing evidence has suggested that the central velocity dispersion of a galaxy is a good predictor of galaxy properties, including star-formation rate (SFR) and colour across large redshift ranges. Furthermore, it is thought to be a more stable quantity with redshift compared to, for example, stellar mass \citep{Bezanson2012,Wake2012}, partly due to the weak dependence of velocity dispersion on both stellar mass and galaxy size, with $\sigma \propto (M_\star / R_e)^{\frac{1}{2}}$. \citet{Leja2013} briefly investigated the change in velocity dispersion for descendants of $z=3$ galaxies selected at a constant cumulative number density. They found a small change ($<0.15$ dex in $\log{\sigma}$) in the average inferred velocity dispersion from $0 < z < 3$ in the G11 SAM. Similarly, simulations of massive galaxies ($\log M_\star > 10.8$) presented in \citet{Oser2012} find an increase in velocity dispersion of $\Delta\sigma = 0.2$ dex over the redshift range from $z=2$ to $z=0$, consistent with observational estimates \citep[e.g.][]{JavierCenarro2009,Martinez-Manso2011}. These observations warrant a detailed study into the use of inferred velocity dispersion in place of stellar mass.

To generate an accurate framework for how galaxies form and evolve, one must observe the evolution of their properties over time. With these observations, models can be crafted to explain them. If, however, the evolution is not traced correctly, these frameworks can deviate from the truth. Although the aforementioned literature works provide some arguments to support the use of their selection methods, no study has attempted to quantify the recovery efficiency of these methods. To this end, we study in detail the ability of these selection methods to trace individual galaxies, as well as their stellar mass and star-formation properties over cosmic time.

In this paper we use models of galaxy evolution to investigate the ability of galaxy selections which are a) above a constant stellar mass limit; b) at a constant cumulative comoving number density in stellar mass; c) above a constant stellar velocity dispersion limit; and d) at a constant cumulative comoving number density in stellar velocity dispersion. We test how well these methods trace the true evolution of progenitor and descendant populations initially selected at redshifts $z=0$ and $z=3$ respectively.

This paper is constructed in the following way. In $\S2$ we discuss the data and the metrics we use to test the efficacy of the observational selection methods. In $\S3$ we present the results and analysis of our work when selecting galaxies using their stellar mass. In $\S4$ we present results when selecting galaxies using their inferred stellar velocity dispersions. Finally, we discuss and conclude our main results in $\S5$ and $\S6$. We describe the simulations and associated cosmologies we use in section \S\ref{sec:simulated-data}.

\section{Data \& Selection Methods}

In this paper, we determine the efficacy of two different methods in recovering the direct (i.e. most massive) progenitors and descendants of the most massive galaxies from initial selections at redshifts of $z = 0$ and $z = 3$, respectively. The first selection method is at a constant limit (in either stellar mass or velocity dispersion), above which galaxies are selected. The second selection is at a constant cumulative comoving number density. This is achieved by integrating the galaxy stellar mass function (GSMF), or the galaxy velocity dispersion function (GVDF) if velocity dispersion is used, to obtain the integrated number density as a function of stellar mass (or velocity dispersion). From this we obtain the stellar mass limit above which all the galaxies are below a certain number density. The sample we examine at that redshift thus contains all galaxies with a stellar mass greater than this value. Additionally, we want to quantify how well each selection method recovers both the average and sum total stellar mass in the descendant or progenitor populations, as well as the average SFR. The combination of cosmological dark matter simulations and semi-analytical recipes, as well as cosmological hydrodynamical simulations, continue to provide the only environments in which to conduct such an investigation.

\subsection{Simulated Data}
\label{sec:simulated-data}

To test this, we utilise the output of the Millennium Simulation (MS) and the catalogues of four SAMs applied to it and its variants. The original simulation consists of $2160^3$ dark matter particles of mass $8.6 \times 10^8\ \textrm{h}^{-1}\ \textrm{M}_{\odot}$ within a comoving box of size $500\ \textrm{h}^{-1}\ \textrm{Mpc}$ on a side. The MS uses a $\Lambda$CDM cosmological model with $\Omega_\mathrm{m} = 0.25$, $\Omega_{\mathrm{b}} = 0.045$, $\mathrm{h} = 0.73$, $\Omega_{\Lambda} = 0.75$, $\mathrm{n}_{\mathrm{s}} = 1$ and $\sigma_{8} = 0.9$, where the Hubble constant is parametrized as $\mathrm{H}_0 = 100\ \mathrm{h\ km\ s}^{-1}\ \mathrm{Mpc}^{-1}$. The \citet[][G13]{Guo2013} SAM utilises a subsequent simulation which follows $2160^3$ particles of mass $9.3 \times 10^8\ \textrm{h}^{-1}\ \textrm{M}_{\odot}$ using an updated WMAP7 \citep{Komatsu2011} cosmology with $\Omega_\mathrm{m} = 0.272$, $\Omega_{\mathrm{b}} = 0.0455$, $\mathrm{h} = 0.704$, $\Omega_{\Lambda} = 0.728$, $\mathrm{n}_{\mathrm{s}} = 0.967$ and $\sigma_{8} = 0.81$. The main difference between these two simulations is the value of the linear power spectrum amplitude on scales of $8\ \textrm{h}^{-1}$\ Mpc, $\sigma_{8}$, which affects the dark matter halo merger rate, and thus the galaxy merger rate, inside the simulation \citep{Conselice2014}.

\citet[][B06]{Bower2006} presents an updated variant of the Durham semi-analytical model of galaxy formation \citep{Cole2000} in which the treatment of active galactic nuclei (AGN) and stellar feedback on halo quenching is improved. They find that these updated treatments reduce the number densities of higher mass galaxies and remove cooling flows from rich clusters. \citet[][D06]{DeLucia2006} applied their model to the output of the MS with updated treatments for stellar populations, dust attenuation and cooling flow suppression via AGN feedback. They find that supernovae and AGN feedback processes play a vital role in the early quenching of star-formation in the progenitors of local brightest cluster galaxies (BCGs). G11 describe an updated model of galaxy formation and evolution with new recipes for supernovae feedback and galaxy bulge sizes among others. They find that the simulated abundance of massive galaxies, with $\log M_\star\ [M_\odot] > 11.0$, are consistent with observations out to $z\sim 1.0$. However, they over predict galaxies of lower stellar mass beyond $z\sim0.6$ and under predict massive galaxies at $z > 1.0$ by at least an order of magnitude (see Fig. 23 in G11). Finally, G13 describe the results of implementing their SAM in a WMAP7 cosmology. They find a requirement for weaker feedback and star-formation efficiency than a WMAP1 cosmology in order to reproduce the observed local GSMF. Merger trees used by B06 are described in \citet{Harker2006} while those employed by the remaining SAMs are presented in \citet{Springel2005}. It is these models based upon these merger trees extracted from the Millennium Simulation from which we study the observational selection methods.

Physically motivated semi-analytical models of galaxy formation \citep[see, for example, ][]{Bower2006,Vogelsberger2014} applied to cosmological dark matter simulations, such as the MS, provide an unparalleled tool to probe the evolution of dark matter halos and the galaxies that reside within them. Simulations and observations at low redshift ($z < 2$) are found to be consistent in many respects (e.g. luminosity functions), however SAMs have varying degrees of success in matching observational quantities beyond this. Comparison of different SAMs and other models show simulated galaxy stellar mass functions are generally consistent with most observations out to $z\sim2$ if feedback mechanisms from AGN and supernovae are included \citep{Croton2006,Lu2013}. This agreement extends to comparison between simulated and observed major mergers for the most massive galaxies inside the MS \citep{Bertone2009}, but not for lower mass systems.

\subsection{Selection Method Metrics}

It is prudent to measure how accurately each selection method samples the progenitors or descendants of the galaxy population being studied. In this work, descendants of an initial $z=3$ selection are identified by following the `descendantId' property in the SAM output catalogues. At each step, duplicate descendants (due to mergers between two or more galaxies) are removed such that the number of true descendants decreases with time. On the other hand, progenitors of an initial $z=0$ selection are defined as the most massive galaxy in the previous redshift snapshot that came to be the galaxy in the current snapshot. These definitions allow traversing of different branches along merger trees, depending on the direction we take. In this work we measure the ability of each selection method to recover galaxy properties using various metrics.

Firstly, the recovery fraction quantifies how many of the available progenitors or descendants are recovered at different redshifts in the sample obtained using a given selection method such that
\begin{equation}
\label{eqn:frec}
f_{\mathrm{rec}} = N_s\ /\ N_{\mathrm{tot}},    
\end{equation}
\noindent where $N_s$ is the number of descendants/progenitors included in the observational selection, and $N_{\mathrm{tot}}$ is the total number of descendants/progenitors available to be selected.

Although helpful, it may not strictly be necessary to sample the descendants or progenitors of interest to reproduce the true evolution of galaxy properties - sampling different galaxies from the true progenitors or descendants might be sufficient if the galaxies replacing those lost have similar properties. Therefore a low recovery fraction may not necessarily correspond to an inability to recover the true evolution in, for example, average stellar mass or SFR. Because of this, we also consider the fraction of the observed sample that is not a galaxy of interest, and call this the contamination fraction, defined as 
	\begin{equation}
    	f_{\mathrm{contam}} = (N_{\mathrm{sel}} - N_s)\ /\ N_{\mathrm{sel}}
    \label{eqn:fcontam}
    \end{equation}
\noindent where $N_{\mathrm{sel}}$ is the number of galaxies within the observed selection and $N_s$ is the number of true (i.e. most massive) descendants/progenitors included in the selection. In this paper we also compare the true mean stellar mass, $\widetilde{m}^\star_{\mathrm{true}}$, of the progenitors or descendants to that observed using each selection method, $\widetilde{m}^\star_{\mathrm{obs}}$, defined as 
      \begin{equation}
      \kappa_{m^\star} = (\widetilde{m}^\star_{\mathrm{obs}} - \widetilde{m}^\star_{\mathrm{true}}) / \widetilde{m}^\star_{\mathrm{true}} = \Delta \widetilde{m}^\star / \widetilde{m}^\star_{\mathrm{true}}.
      \end{equation}
In a similar fashion, the ability to trace the evolution of the stellar mass density, or sum of the stellar mass, is important to understand the build up of stellar mass in galaxies over time. We compare the observational selection techniques' abilities to return the true mean stellar mass density, quantified as
      \begin{equation}
      \kappa_{\rho^\star} = \Delta (\Sigma \ {m^\star}) / \Sigma \ {m^\star_{\mathrm{true}}}
      \label{eqn:frho}
      \end{equation}
      
\noindent where $(\Sigma \ {m^\star})$ is the sum of stellar masses. Finally, we extend this to the average SFR of the galaxies in a similar fashion such that the discrepancy between the true and observed is defined as

      \begin{equation}
      \kappa_{\Psi^\star} = \Delta \widetilde{\Psi}^\star \ / \ \widetilde{\Psi}^\star_{\mathrm{true}}
      \end{equation}

\noindent where $\widetilde{\Psi}^\star$ is the mean SFR. We choose to study these galaxy properties in particular because they are the most fundamental, and the most used in the literature thus far. Additionally, we investigate whether each selection method is best applied to tracing progenitor or descendant galaxy populations, i.e. whether the selection methods are best applied forwards or backwards in time.

\subsection{Velocity Dispersion Selection}

As central velocity dispersion has been shown to exhibit a shallow evolution over time, it is prudent to investigate this physical property as a tracer. From scalar virial theory the stellar velocity dispersion of a system can be estimated by

\begin{equation}
\label{eqn:sigma}
	\sigma^2 = \frac{GM_\star}{5R_{\frac{1}{2}}}
\end{equation}

\noindent where $G$ is Newton's gravitational constant, $M_\star$ is the total stellar mass and $R$ is the half-mass radius (see, e.g., \citealp{Cappellari2006}). Using the reported total and bulge stellar masses and sizes, we follow \citet{Leja2013} in estimating the half-mass radius of each simulated galaxy as 

\begin{equation}
	\label{eqn:galsize}
    R_{\frac{1}{2}} = \frac{M_b R_b + M_d R_d}{M_b + M_d}.
\end{equation}

\noindent As before, $M$ is stellar mass, $R$ is the half-mass radius and subscripts $b$ and $d$ correspond to the bulge and disk components respectively. We correct the disk scale radius provided in the SAM catalogues to convert it to a half-mass radius such that $R_d = 1.678 R_{\mathrm{scale}}$. This relation is ideally obtained numerically, however approximations are available for a range of \citet{Sersic1963} indices . We refer the interested reader to \citet{Graham2005} and references therein for more information.

We calculate these metrics using both stellar mass and velocity dispersion at four constant number density selections and four constant limits. Number density values are chosen to cover a wide range, allowing comparison with previous work, and to be representative of what number densities are currently applicable to observational studies. Constant limits in both stellar mass and velocity dispersion are chosen to enable comparison with the number density selections. Constant stellar mass limits are defined as the mass limit of a number density selection at either $z=0$ or $z=3$, depending on whether progenitors or descendants are being investigated. In short, the limits are chosen such that the initial selections, whether at $z=0$ or $z=3$, are the same. In \S\ref{sec:stellar-mass-results} we report our findings using stellar mass as the ranking property and in \S\ref{sec:sigma-results} we report the results of using inferred velocity dispersion.

\section{Stellar Mass Selections}
\label{sec:stellar-mass-results}

We investigate the ability of two different galaxy selection methods, using two different galaxy properties, to recover the mean stellar mass, stellar mass density, mean velocity dispersion and average SFR of progenitor and descendant populations. All our results are available to download online\footnote{\url{http://goo.gl/X6QZGC}}.

\begin{figure}
\center{\includegraphics[width=\columnwidth]{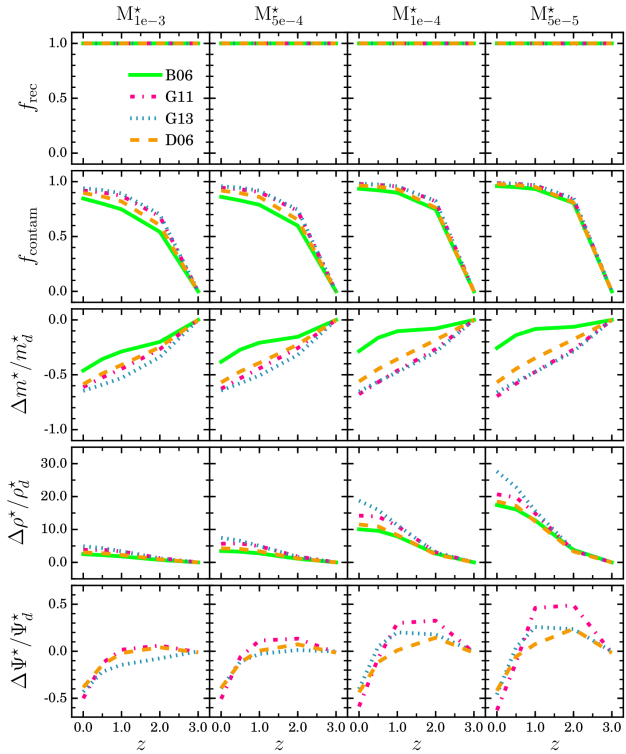}}
\caption{Recovery fraction of individual descendants, their average stellar mass, stellar mass density and average SFR (rows) for four constant stellar mass limit selections (columns) covering the stellar mass range at $z=3$ of $\log\ M_* > 10.7, 10.6, 10.4, 10.2$. Initial stellar mass selection limits for each SAM are given in the third column of Table \ref{tab:mass-lims}. SAMs used are \citet[][B06]{Bower2006}, \citet[][D06]{DeLucia2006}, \citet[][G11]{Guo2011} and \citet[][G13]{Guo2013}, represented by solid green, dashed orange, dashed-dotted magenta and dotted blue lines respectively. (A colour version of this figure is available in the online journal.)}
\label{fig:desc-stellar-c}
\end{figure}

\begin{figure}
\center{\includegraphics[width=\columnwidth]{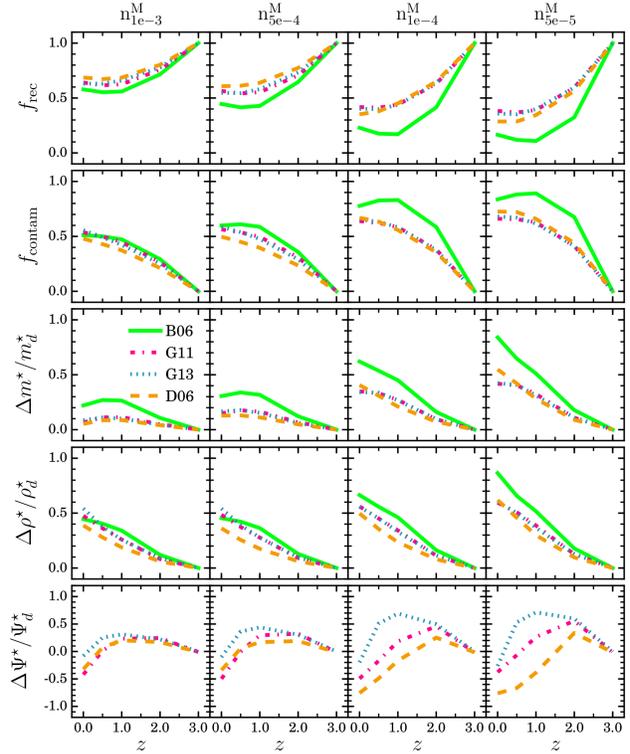}}
\caption{Recovery fraction of individual descendants, their average stellar mass, stellar mass density and average SFR (rows) for four constant number density selections (columns) covering the number density range $-4.3 < \log\ n < -3.0$. These approximately correspond to stellar masses at $z=3$ of $\log\ M_* > 10.7, 10.6, 10.4, 10.2$. Initial stellar mass selection limits for each SAM are given in the third column of Table \ref{tab:mass-lims}. SAMs used are \citet[][B06]{Bower2006}, \citet[][D06]{DeLucia2006}, \citet[][G11]{Guo2011} and \citet[][G13]{Guo2013}, represented by solid green, dashed orange, dashed-dotted magenta and dotted blue lines respectively. (A colour version of this figure is available in the online journal.)}
\label{fig:desc-stellar-n}
\end{figure}

\subsection{Descendants}
\label{sec:stellar-desc}
In Figure \ref{fig:desc-stellar-c} we show the results of selecting descendants at four constant stellar mass limits, described in Table \ref{tab:mass-lims}. This is defined as the stellar mass limit for a number density selection beginning at $z=3$ and examining evolution at lower redshift. As one might expect, all constant stellar mass selections recovered all available descendant galaxies at every redshift (top panel), as galaxies typically experience a net gain in stellar mass over these redshift ranges. However, the fraction of the selected sample that are not descendants of interest, $f_{\mathrm{contam}}$, increases to $\sim50\%$ by $z=2$. By $z=0$, samples selected above each stellar mass limit are almost completely contamination. This shows that using a constant mass cut at $z=3$ selects essentially none of the same galaxies (descendants) at lower redshift.

Contrary to what this metric might suggest, the difference between the observed and true mean stellar mass is underestimated by only $~50\%$ by $z=0$, decreasing linearly at lower redshift. This seems to be largely independent of the stellar mass limit in all but the B06 SAM which fares relatively better at higher limits. Recovery of the median stellar mass is indistinguishable from the mean stellar mass for the two smallest number densities. At the two largest choices however, the median stellar mass is further underestimated towards lower redshift such that at $z=0$, this property is underestimated by $\sim 60\%$. The stellar mass density is overestimated by a factor of $\sim 4$ ($\sim 20$) at the lowest (highest) mass limit in all SAMs by $z=0$. As the B06 SAM does not report SFRs in its catalogues, we instead consider only the remaining three SAMs in recovery of the mean SFR. At the lowest stellar mass limit, this is recovered to within $\sim10\%$ down to $z=1$. At lower redshifts, however, the SFR begins to be increasingly underestimated and by $z=0$ it is underestimated by $\sim50\%$.

We show in Figure \ref{fig:desc-stellar-n} that by using a constant number density selection the recovery fraction, $f_{\mathrm{rec}}$, decreases exponentially with decreasing redshift such that by $z=0$ between $30\%$ and $60\%$ of the available descendants are selected. The contamination fraction is found to vary between half and three quarters of the sample at the highest and lowest number densities respectively. Mean stellar mass is overestimated by a factor that increases with both number density and redshift, overestimating the true value by a factor of $1.3$ and $1.6$ by $z=0$. While not plotted, the median stellar mass is qualitatively similar above $z \sim 0.5$. Below this, the median stellar mass is further overestimated by up to $30\%$ more at $z=0$. Similarly, the stellar mass density is eventually overestimated by a factor of $1.5$ ($1.8$) at the largest (smallest) number densities. Finally, over the entire redshift range the SFR is recovered to within $\sim50\%$ in all SAMs and at all number density choices. 

Compared to a constant stellar mass selection, a constant number density selection recovers far fewer of the true descendants at lower redshift. However, at all number density choices, the lower redshift selections have considerably less contamination. 

\begin{table}
\caption{Constant stellar mass limits for progenitors and descendants, defined as the stellar mass limit for a number density selection at $z=0$ and $z=3$ respectively.}
\label{tab:mass-lims}
\begin{tabular}{@{}cccc}
	\hline \hline
	\multirow{2}{*}{SAM} & \multirow{2}{*}{$n$ [Mpc$^{-3}\ h^3$]}	& Descendants ($z=3$) & Progenitors ($z=0$) \\
    & & $M_{\mathrm{lim}}$ [$\log M_{\odot} / h$] &	$M_{\mathrm{lim}}$ [$\log M_{\odot} / h$]\\
    \hline
    B06 & $1\times10^{-3}$ & 10.32 & 10.87\\
    & $5\times10^{-4}$ & 10.51 & 11.01 \\
    & $1\times10^{-4}$ & 10.76 & 11.22 \\
    & $5\times10^{-5}$ & 10.82 & 11.30 \\
    \hline
    D06 & $1\times10^{-3}$ & 10.31 & 10.92 \\
    & $5\times10^{-4}$ & 10.46 & 11.04 \\
    & $1\times10^{-4}$ & 10.68 & 11.29 \\
    & $5\times10^{-5}$ & 10.74 & 11.39 \\
	\hline
    G11 & $1\times10^{-3}$ & 10.21 & 10.84\\
    & $5\times10^{-4}$ & 10.33 & 10.97 \\
    & $1\times10^{-4}$ & 10.52 & 11.24 \\
    & $5\times10^{-5}$ & 10.59 & 11.35 \\
    \hline
    G13 & $1\times10^{-3}$ & 10.00 & 10.77 \\
    & $5\times10^{-4}$ & 10.15 & 10.89 \\
    & $1\times10^{-4}$ & 10.40 & 11.12 \\
    & $5\times10^{-5}$ & 10.48 & 11.21 \\
    \hline \hline
\end{tabular}
\end{table}

\subsection{Progenitors}

\begin{figure}
\center{\includegraphics[width=\columnwidth]{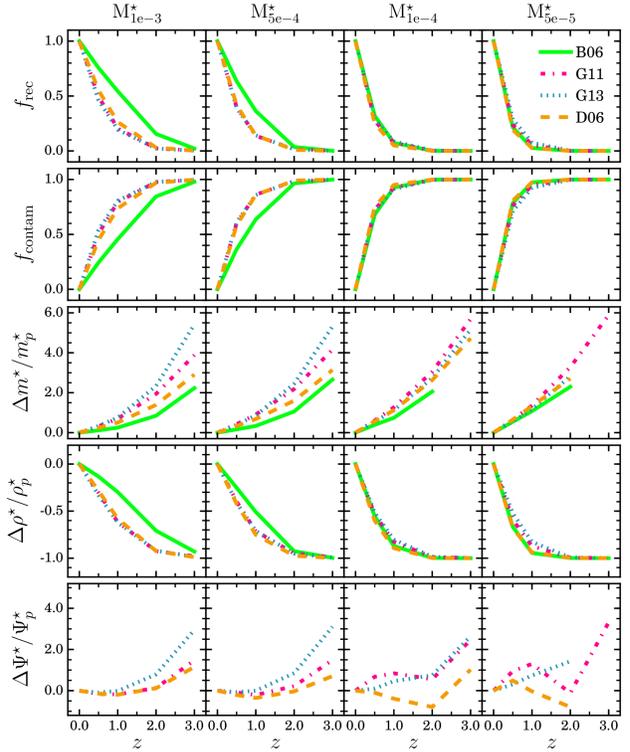}}
\caption{Recovery of individual progenitors, their average stellar mass, stellar mass density and average SFR (rows) for four constant stellar mass limit selections (columns) covering the number density range $-4.3 < \log\ n < -3.0$. The B06, D06, G11 and G13 models represented by solid green, dashed orange, dashed-dotted magenta and dotted blue lines respectively. Progenitor stellar mass limits for each SAM are given in the third column of Table \ref{tab:mass-lims}. (A colour version of this figure is available in the online journal.)}
\label{fig:prog-stellar-c}
\end{figure}

Now we take an initial selection at $z=0$ and trace the most massive progenitors back to higher redshifts. As Figure \ref{fig:prog-stellar-c} shows, the recovery of individual progenitors using a constant stellar mass limit (detailed in Table \ref{tab:mass-lims}) deteriorates exponentially such that at $z=1$, only $30\%$ are recovered in the selection at the smallest stellar mass limit and less than $5\%$ at the largest stellar mass limit (smallest number density). The sample's contamination fraction increases immediately and, at all stellar mass limits, the sample is $>95\%$ contamination by $z=3$.

The discrepancy between the true and observed mean stellar mass of the progenitors increases approximately linearly with redshift, overestimating the mean mass by a factor of three at $z=3$, independent of stellar mass limit and weakly dependent on the choice of SAM. Recovery of the median stellar mass is again indistinguishable from the mean recovery at the two largest stellar mass limits. At the two highest, the median mass is overestimated by factors of $3-7$ times. Furthermore, observed stellar mass density is increasingly underestimated with redshift in all SAMs. Finally, the mean SFR is recovered to within a factor of $\sim4$ by $z=3$.

\begin{figure}
\center{\includegraphics[width=\columnwidth]{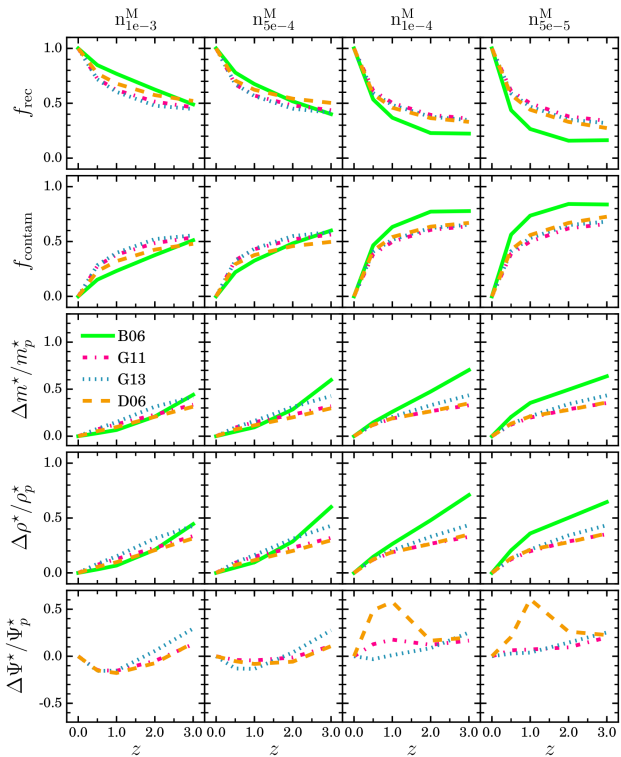}}
\caption{Recovery of individual progenitors, their average stellar mass, stellar mass density and average SFR (rows) for four constant number density selections (columns) covering the number density range $-4.3 < \log\ n < -3.0$. The B06, D06, G11 and G13 models represented by solid green, dashed orange, dashed-dotted magenta and dotted blue lines respectively. (A colour version of this figure is available in the online journal.)}
\label{fig:prog-stellar-n}
\end{figure}

Selecting progenitors at a constant number density fares relatively better, as shown in Figure \ref{fig:prog-stellar-n}. Out to $z=3$, no less than $\sim50\%$ ($\sim30\%$) of progenitors are recovered at the largest (smallest) selections. The mean stellar mass is recovered to within a factor of $\sim1.5$ at $z=3$ in all SAMs and choices of $n$ and the observed stellar mass density follows a very similar trend. Median stellar mass recovery is indistinguishable from the mean recovery except at the largest number density. Here up until $z \sim 2$ where the overestimation becomes larger by approximately $\sim 50\%$. Lastly, the mean SFR is recovered to within $\pm20\%$ at all number density selections except for D06, which overestimates the SFR at a peak of $\sim50\%$ at $z=1$.

Comparing these results with the descendant population (\S\ref{sec:stellar-desc}), we find that a constant cumulative comoving number density selection recovers all descendent population properties within a factor of two of the true value. Similarly, all the progenitor properties are recovered to within a factor of 1.5 of the true value. Furthermore, a constant number density selection appears to trace the ensemble progenitor properties of $z=0$ massive galaxies better than the descendants of those at $z=3$.

\subsection{Fitting Forms}
\label{sec:fitting}

For convenience, we fit parametric functions to the metrics described above for a constant number density selected sample of galaxies. These fits are strictly valid over the redshift range $0 < z < 3$. For each metric, we take the average value of all SAMs at each redshift. Next, at each redshift we randomly sample a Gaussian of width equal to the spread between SAMs, centred on the mean value. The function is then fit to these points. These steps are repeated $10^4$ times to obtain the average parameters and their associated errors for each metric at each number density. Our detailed results of fitting for all number densities are reported in Table \ref{tab:fitting-results-n}.

The descendant galaxy recovery and contamination fraction is fit with a function of form
\begin{equation}
\label{eqn:recfracformfit}
	f = a + b \times \exp\left({c \times (1+z)}\right),
\end{equation}
\noindent where $a$, $b$ and $c$ are free parameters. We find that forcing $a$ to zero for the recovery fraction, and $c$ to unity for the contamination fraction, gives better fits. The recovery of average stellar mass ($\kappa_{m^\star}$) and stellar mass density ($\kappa_{\rho^\star}$) are then parametrised as
\begin{equation}
\label{eqn:massformfit}
	\kappa = a + b \times (1+z) .
\end{equation}

\noindent To fit the progenitor galaxy contamination fraction, we modify the parametrisation to the form
\begin{equation}
\label{eqn:contamprogformfit}
	f' = a + b \times (1+z) + (1+z)^c .
\end{equation}

Finally, we do not fit the recovery of average SFR recovery. Fitted parameters and associated errors are available in Table \ref{tab:fitting-results-n} for constant cumulative number density selected samples.

\section{Inferred Velocity Dispersion Selections}
\label{sec:sigma-results}

In an era of ever larger and deeper spectroscopic surveys, the extra information these observations afford of internal properties could possibly be employed as a better tracer of progenitor or descendant galaxy properties. One product of such a survey is the measurement of the central stellar velocity dispersion of a galaxy.

As the SAM catalogues do not report the velocity dispersions of galaxies, we infer this quantity using Equations \ref{eqn:sigma} and \ref{eqn:galsize}. The inferred velocity dispersion, as defined here, is a relatively direct observable at the redshifts probed because the stellar mass and half-mass radius are observable. In this section we report the results of using this property in place of stellar mass. As a reminder, this would be achieved observationally by integrating the GVDF to obtain the cumulative number density of galaxies as a function of their velocity dispersion. The velocity dispersion limit, above which all galaxies are at a number density $n$, can simply be read off. 

\subsection{Descendants}

\begin{figure}
\center{\includegraphics[width=\columnwidth]{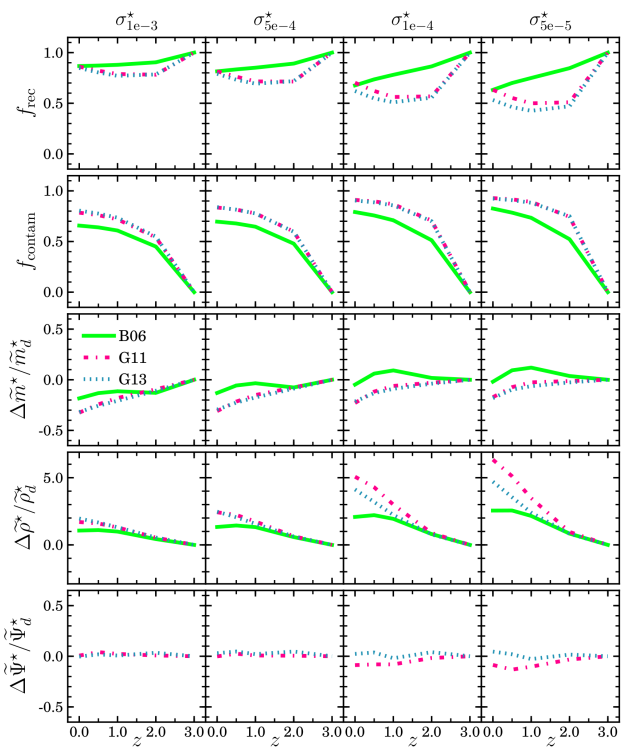}}
\caption{Recovery of individual progenitors, their average stellar mass, stellar mass density and average SFR (rows) at four constant inferred velocity dispersion selections (columns) covering the range $-4.3 < \log\ n < -3.0$. The B06, G11 and G13 models are represented by solid green, dashed-dotted magenta and dotted blue lines respectively. Inferred velocity dispersion limits given in Table \ref{tab:sigma-lims}. (A colour version of this figure is available in the online journal.)}
\label{fig:desc-sigma-c}
\end{figure}

As displayed in Figure \ref{fig:desc-sigma-c}, selecting galaxies above a constant inferred velocity dispersion limit, given in Table \ref{tab:sigma-lims}, results in slowly losing descendants with decreasing redshift in the B06 SAM. However, in the G11 and G13 SAMs, the recovery fraction increases below $z < 2$. At $z=0$, $90\%$ and $60\%$ of descendant galaxies are selected above the lowest and highest inferred velocity dispersion limits respectively. As with selection above a constant stellar mass limit, the contamination fraction increases exponentially towards lower redshift. At the lowest (highest) velocity dispersion limits there is significant contamination in the observed sample at the level of $70\%$ ($90\%$). Recovery of the descendant mean stellar mass is increasingly underestimated. The true value is maximally underestimated at $z=0$ at all inferred velocity dispersion limits by $30\%$. Conversely, the stellar mass density is increasingly overestimated with time by up to a factor of $\sim5$ times the true value. Finally, a constant inferred velocity dispersion selection recovers the descendants' average SFR to within $10\%$ at all limit choices and redshifts.

\begin{figure}
\center{\includegraphics[width=\columnwidth]{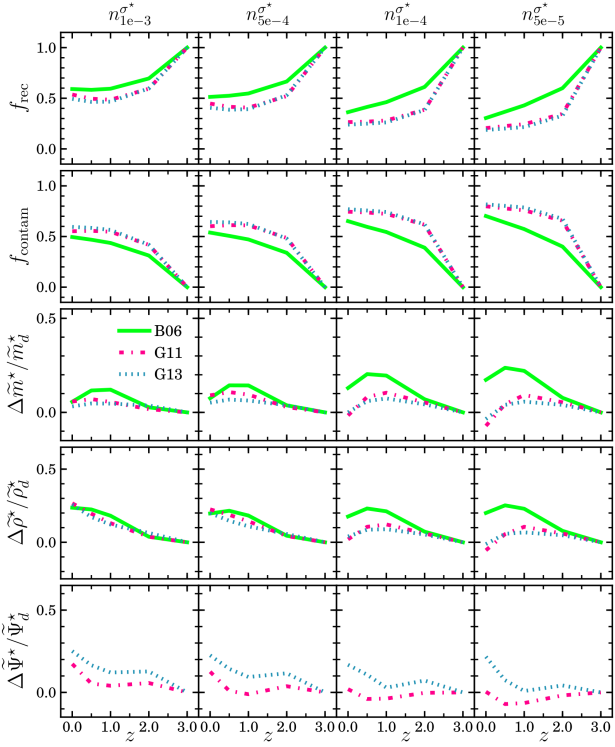}}
\caption{Recovery of individual progenitors, their average stellar mass, stellar mass density and average SFR (rows) for four constant number density selections (columns) covering the range $-4.3 < \log\ n < -3.0$. The B06, G11 and G13 models are represented by solid green, dashed-dotted magenta and dotted blue lines respectively. (A colour version of this figure is available in the online journal.)}
\label{fig:desc-sigma-n}
\end{figure}

In Figure \ref{fig:desc-sigma-n} we show the result of selecting at a constant cumulative number density in inferred velocity dispersion. Inferred velocity dispersion is, as defined in this paper, a function of and proportional to stellar mass for each galaxy type (early and late) and so it is not surprising that the results are similar to those obtained in \S\ref{sec:stellar-mass-results}. Comparing with the stellar mass selections described in \S\ref{sec:stellar-desc}, these results suggest inferred velocity dispersion is just as competent a tracer as stellar mass, and even more accurate in some cases. However, any improvements are small over the use of stellar mass. 

\subsection{Progenitors}

\begin{figure}
\center{\includegraphics[width=\columnwidth]{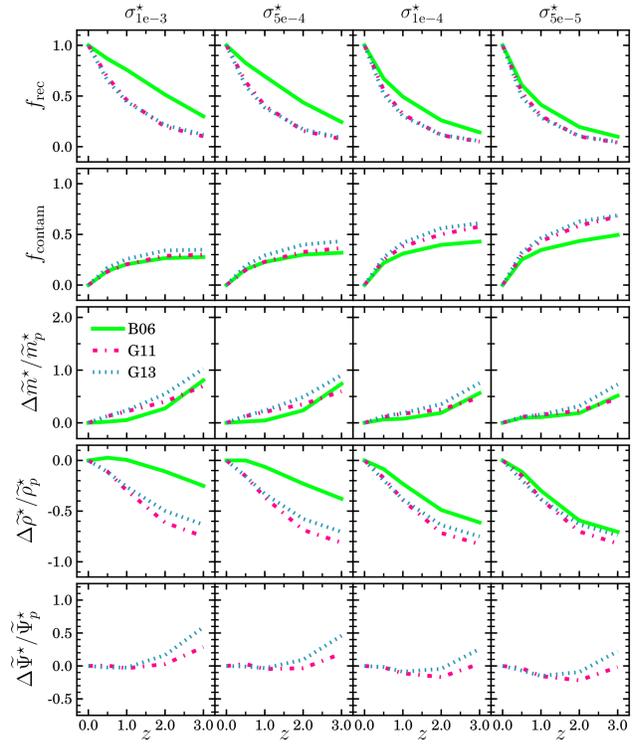}}
\caption{Recovery of individual progenitors, their average stellar mass, stellar mass density and average SFR (rows) at four constant inferred velocity dispersion selections (columns) covering the range $-4.3 < \log\ n < -3.0$. The B06, G11 and G13 models are represented by solid green, dashed-dotted magenta and dotted blue lines respectively. Inferred velocity dispersion limits given in Table \ref{tab:sigma-lims}. (A colour version of this figure is available in the online journal.)}
\label{fig:prog-sigma-c}
\end{figure}

\begin{figure}
\center{\includegraphics[width=\columnwidth]{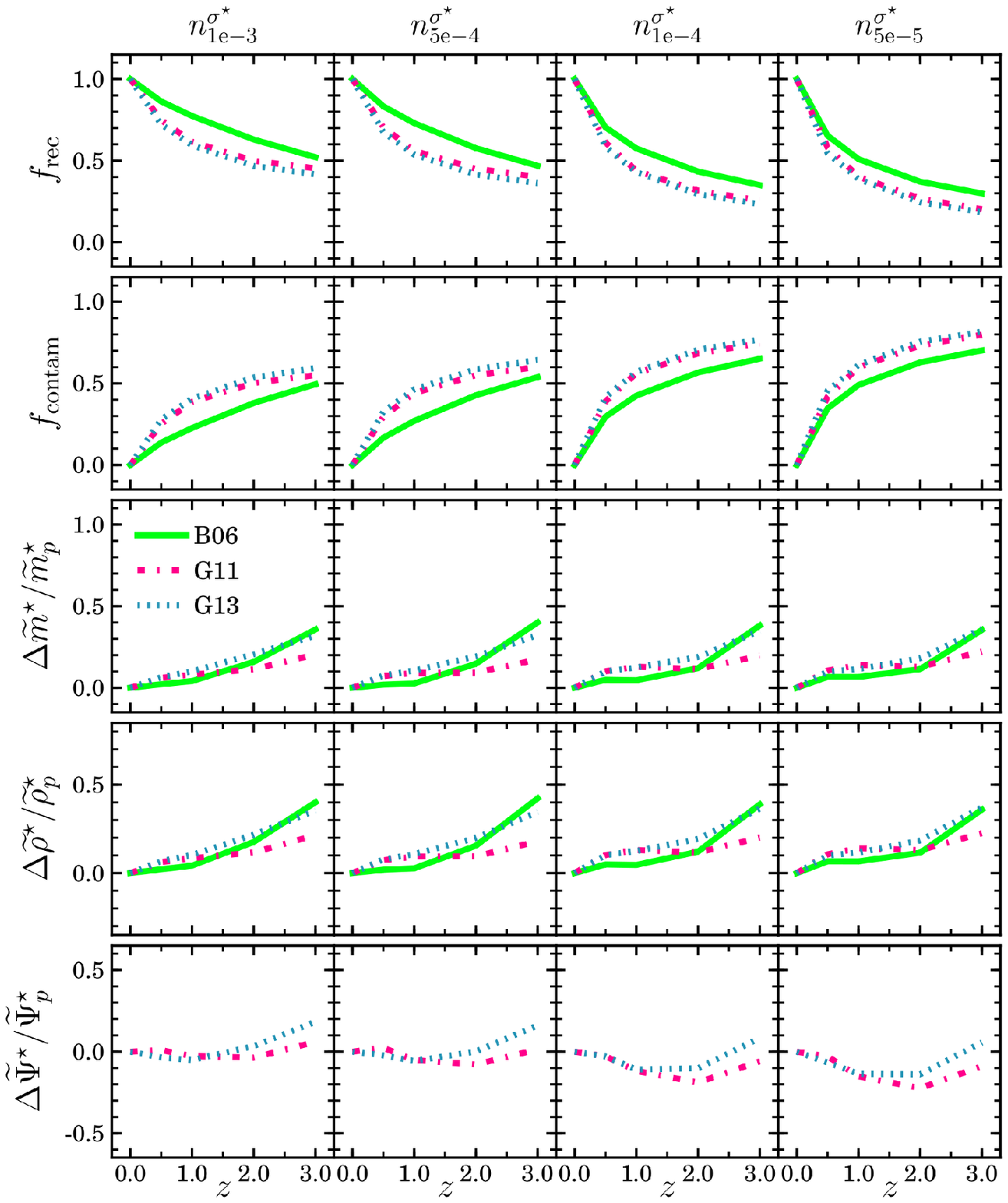}}
\caption{Recovery of individual progenitors, their average stellar mass, stellar mass density and average SFR (rows) for four constant number density selections (columns) covering the range $-4.3 < \log\ n < -3.0$. The B06, G11 and G13 models are represented by solid green, dashed-dotted magenta and dotted blue lines respectively. (A colour version of this figure is available in the online journal.)}
\label{fig:prog-sigma-n}
\end{figure}

Figures \ref{fig:prog-sigma-c} and \ref{fig:prog-sigma-n} display the results of attempting to trace the progenitors of $z=0$ galaxies via selection above a constant inferred velocity dispersion and at a constant cumulative number density, ordered by inferred velocity dispersion, respectively. Selection above a constant limit loses progenitor galaxies from the sample with increasing redshift. By $z=3$, only $10\%$ of true progenitors are sampled at all velocity dispersion limits and SAMs. The average stellar mass is increasingly overestimated with redshift, by $50-100\%$ at $z=3$. The stellar mass density is increasingly underestimated. In the B06 SAM, it is even more underestimated at higher inferred velocity dispersion limits than at smaller limits. However, in the G11 and G13 SAMs, it is underestimated by $70\%$ at all limits by $z=3$. Similarly, the SFR is recovered to within $50\%$ in the two SAMs considered.

Selection at a constant cumulative number density, ordered by inferred velocity dispersion, results in recovery fractions similar to the number density selection using stellar mass. Decreasing exponentially from $z=0$, $50\%$ ($30\%$) at the largest (smallest) number densities. Both the average stellar mass and the stellar mass density are recovered to within $40\%$ of the true value across all redshifts, SAMs and number densities investigated in this paper. Finally, the SFR is recovered to within $20\%$ at all times.

\begin{table}
\caption{Inferred velocity dispersion limits for progenitors and descendants, defined as the minimum inferred velocity dispersion for a number density selection at $z=0$ and $z=3$ respectively. Inferred velocity dispersions are calculated using Equation \ref{eqn:sigma} with the galaxy component stellar masses and sizes reported by each SAM.}
\label{tab:sigma-lims}
\begin{tabular}{@{}cccc}
	\hline \hline
	\multirow{2}{*}{SAM} & \multirow{2}{*}{$n$ [Mpc$^{-3}\ h^3$]}	& Descendants & Progenitors \\
    & & $\sigma_{\mathrm{lim}}$ [$\mathrm{km}\ \mathrm{s}^{-1}$] &	$\sigma_{\mathrm{lim}}$ [$\mathrm{km}\ \mathrm{s}^{-1}$]\\
    \hline
    B06 & $1\times10^{-3}$ & 133.1 & 198.2\\
    & $5\times10^{-4}$ & 181.7 & 263.5 \\
    & $1\times10^{-4}$ & 309.3 & 424.5 \\
    & $5\times10^{-5}$ & 365.4 & 502.4 \\
    \hline
    G11 & $1\times10^{-3}$ & 83.7 & 131.8\\
    & $5\times10^{-4}$ & 100.2 & 156.9 \\
    & $1\times10^{-4}$ & 142.1 & 208.9 \\
    & $5\times10^{-5}$ & 161.6 & 231.0 \\
    \hline
    G13 & $1\times10^{-3}$ & 69.3 & 109.6 \\
    & $5\times10^{-4}$ & 84.7 & 131.8 \\
    & $1\times10^{-4}$ & 124.7 & 181.1 \\
    & $5\times10^{-5}$ & 143.8 & 202.6 \\
    \hline \hline
\end{tabular}
\end{table}

\section{Discussion}

Firstly, let us contrast the use of a constant stellar mass selected sample and a constant number density (in stellar mass) selected sample. As one may have expected, the former recovers all descendants of an initial high redshift sample. This is due to our definition of a descendant and that the stellar mass of systems can only increase with time inside these simulations. Even though the recovery fraction is high, the contamination fraction increases to $>80\%$ within $\sim 2$ Gyr as galaxies, initially unsampled, increase their stellar mass and move into the selection. 

Comparing the recovery and contamination fractions obtained through constant number density selections of descendants and progenitors, one can infer how these populations have evolved. Taking the smallest number density choice of $n = 5 \times 10^{-5}$ Mpc$^{-3}$ $h^3$, at $z=0$ we recover $30\%$ of the available descendants and nearly three quarters of our selection is contamination. Similarly, at $z=3$ we recover $30\%$ of the progenitors and $70\%$ of the sample is contamination. These results suggest that a large fraction of the progenitors of the most massive local galaxies are not the most massive at higher redshifts. Conversely, a large fraction of the most massive galaxies at high redshift are not among the most massive at lower redshifts. The one-to-one mapping in stellar mass rank order that this selection method assumes does not occur within these simulations. Furthermore, lower mass systems from below the selection at high redshift increase their stellar mass at a higher rate than those more massive systems and become most of the most massive galaxies in the local Universe.

It is worth noting that all of the SAMs used in this work fail to match observed galaxy stellar mass functions beyond some redshift (typically $z\sim~1.5$) meaning that they also fail to reproduce the observed evolution of certain galaxy populations. A known problem with the original MS is the cosmology used. Use of a larger $\sigma_8$ than currently observed \citep{Komatsu2011,PlanckCollaboration2015} will increase the merger rate and therefore the scatter in the rank order of galaxy stellar mass. Furthermore, this cosmology produces a larger population of quenched galaxies earlier than observed. This requires the SAMs to build up the low mass end of the GSMF at early times in order to match the observed local stellar mass function. See \citet{Leja2013} and \citet{Guo2011} for an in-depth discussion into this and other issues. We would thus expect less scatter in the real Universe, and therefore better recovery of galaxy properties, compared to the results obtained here. Cosmological hydrodynamical simulations, e.g. \citet{Furlong2014a} and \citet{Genel2014}, show closer agreement with observed galaxy stellar mass functions out to high redshift and may offer a better environment in which to conduct these tests.

\subsection{Where are the progenitors of $z=0$ massive galaxies at high redshift?}

\begin{figure*}
\center{\includegraphics[width=2\columnwidth]{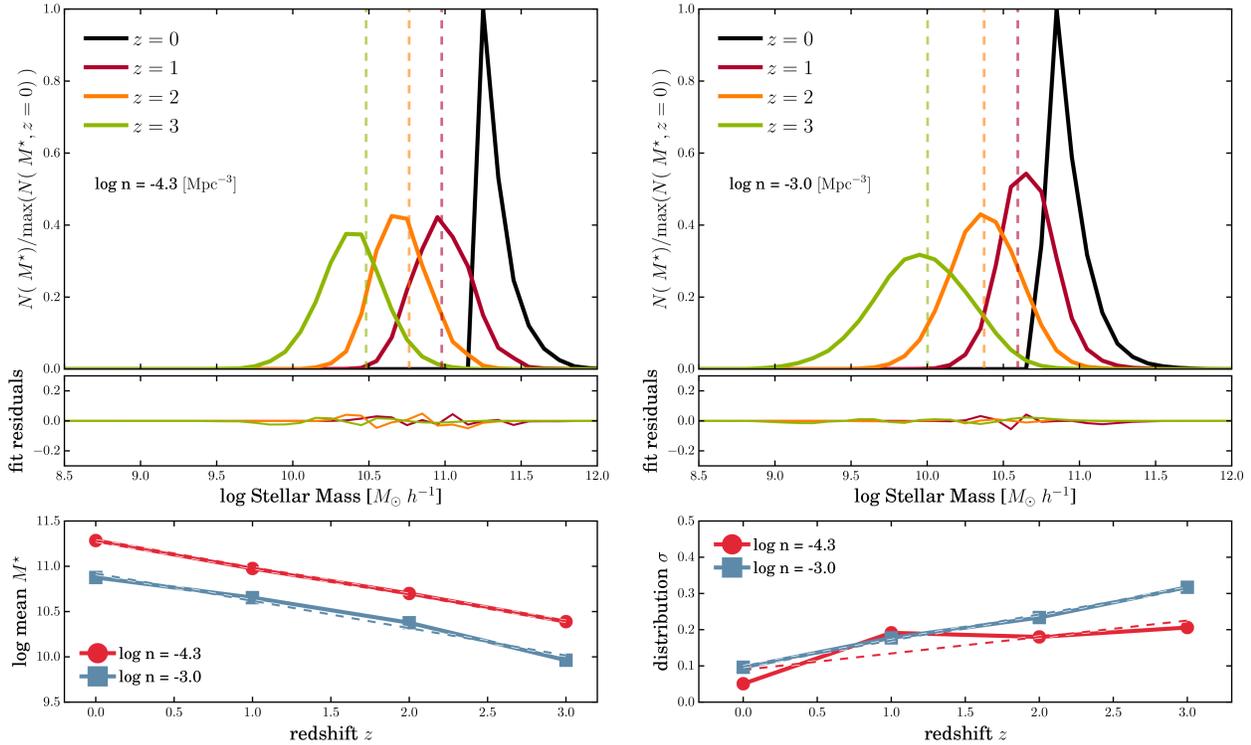}}
\caption{Progenitor mass distributions in the \citet{Guo2013} SAM for $z=0$ galaxy selections at constant number densities $\log\ n = -4.3$ Mpc$^{-3}$ h$^{3}$ (top left) and $\log\ n = -3.0$ Mpc$^{-3}$ h$^{3}$ (top right). These number densities correspond to galaxies with a $z=0$ stellar mass of $\log\ M^\star > 11.21\ [M_\odot \ h^{-1}]$ and $\log\ M^\star > 10.77\ [M_\odot \ h^{-1}]$, respectively. Stellar mass distributions at $z=0,1,2,3$ are given in black, red, orange and green solid lines. Dashed vertical lines represent the stellar mass cuts inferred from a number density cut at each redshift. Residuals from Gaussian fits to these distributions are displayed in the middle panels. Parameters from Gaussian fits to the progenitor masses are shown in the bottom row, with the mean stellar mass and standard deviation at bottom left and bottom right respectively with the largest (blue squares) and smallest (red circles) number densities plotted as a function of redshift. (A colour version of this figure is available in the online journal.)}
\label{fig:prog_mass_dist}
\end{figure*}

A simple question we can ask is where exactly in the ranking (in either stellar velocity dispersion or stellar mass) are the progenitors of the most massive galaxies at $z=0$ at earlier times. Taking the two extremes of the number density choices in this work, we show in Figure \ref{fig:prog_mass_dist} the stellar mass distributions of the most massive progenitors of $z=0$ galaxies in the G13 SAM selected at two extreme number densities (top row). At each redshift we fit the stellar mass distributions with a Gaussian function and show in the middle panels that fit residuals are minimal ($< 5$\%). This is done to quantify the changes in these distributions as a function of redshift. In the bottom row, we show how the properties of these distributions change with redshift. The mean and widths (distribution standard deviation) are shown in the bottom left and bottom right panels respectively.

We find that at higher redshift, the mass distributions move systematically towards smaller mass galaxies and the stellar mass distribution widens. At the highest redshift, the distributions are found to have standard deviations of $\sigma = 0.32$ [$\log M_\odot$ h$^{-1}$] and $\sigma = 0.21$ [$\log M_\odot$ h$^{-1}$] for the largest and smallest number density selections. In both cases, the distributions increase in width by a factor of $\sim 3$ since $z=0$.

Furthermore, the selections made by a constant number density (the mass limits of which are indicated by the vertical dashed lines in the top panels of Figure \ref{fig:prog_mass_dist}) show that beyond $z>1$, the majority of the progenitors are below this limit (i.e. the the peak of the actual progenitor stellar mass distribution is found at a lower stellar mass than the selection's stellar mass limit). Therefore, within the Millennium Simulation at the very least, it can be said the progenitors of the most massive local galaxies are not only the most massive galaxies at higher redshifts - they span a wide range of masses at higher redshifts. For example, the most massive progenitors span more than an order of magnitude ($> 1$ dex) in stellar mass for the largest initial number density selection (top right panel) and $\sim 0.5$ dex for the smallest number density at $z=3$. As shown in this work however, this does not appear to significantly impact the ability of a number density selected sample to recover the average stellar mass, star-formation rate or stellar mass density.

\subsection{How do mergers affect our selections?} 
The number of progenitors or descendant galaxies can change over time due to mergers between objects within the initial sample. Selecting galaxies at a constant number density ignores these changes, and potentially contributes to the over- or underestimation of ensemble properties.

To determine the extent of this, we calculate the number of mergers between the descendants of an initial selection at $z=3$ in the B06 and G13 SAMs, as these are based on different dark matter merger trees from the MS. In B06, $1.3\%$ and $15.5\%$ of galaxies in our initial selection are lost due to mergers from $z=3$ to $z=0$ at the smallest and largest number density selections. In the G13 SAM however, these measurements are higher at $10.9\%$ and $29.6\%$, respectively. For the most massive galaxies, this translates to approximately $3-5$ mergers per massive galaxy (see Table \ref{tab:mass-lims} for mass limits) over the redshift range $0 < z < 3$. It must be noted that these numbers represent all (total) mergers, and are not major mergers as they may include some mergers with mass ratios greater than $1:4$, the most widely used definition. These measurements are slightly higher compared with pair fraction and morphological observations of major mergers in comparably massive galaxies \citep[see, e.g.,][]{Bluck2009, Conselice2009, Man2012}.

At increasingly larger number densities, mergers within the descendant population may become increasingly important. As such, selection at a constant number density may not be applicable over such a redshift range. It may be appropriate to correct the number density between redshift bins to account for mergers that have occurred within the sample. However, the reduction of the number density in response to descendant galaxy mergers does not result in the desired effect. Qualitatively, reducing the number density at each redshift results in higher stellar mass limits. The average stellar mass of the observed sample would therefore increase. As the average stellar mass and stellar mass density are already overestimated, this discrepancy would only increase. On the other hand, the `un-merging' of galaxies going backwards in time would increase the number density used to trace progenitor galaxies. This would lower the stellar mass limit used to select the samples and thus decrease the measured average stellar mass of the observed samples. As this quantity is also overestimated, this discrepancy would be reduced. However, as we have increased the number density contamination would also increase.

\subsection{Can we infer velocity dispersion in a semi-analytical model?}
We investigate inferred velocity dispersion in place of stellar mass as a ranking property due to evidence of a shallower and more stable evolution with redshift. It is prudent to ask whether this quantity can be accurately obtained from the SAMs used in this work. Use of Equation \ref{eqn:sigma} implicitly assumes a spherically symmetric system and would correspond to a system with a S\'ersic index of $n \approx 5.5$ \citep{Cappellari2006}. However, the factor in the denominator doesn't account for multiple components (i.e. a bulge and disk) and is influenced not only by S\'ersic index but also galaxy black hole mass. Thus the value calculated using this equation is a simplistic estimate at best and not strictly applicable to every type of galaxy. Furthermore, disk-dominated systems are not spherical and isotropic and thus this equation is not strictly applicable to these types of systems.

Using the bulge-to-total stellar mass ratio (B/T) as a proxy for disk and bulge dominated morphologies, we find that the most massive galaxies at $z = 0$ ($z=3$) in the B06 SAM are typically bulge dominated with only $30\%$ ($40\%$) having B/T $< 0.5$. While this suggests Equation \ref{eqn:sigma} is applicable at these redshifts, this SAM does not reproduce observations of larger disk-dominated fractions at high redshift \citep{Bluck2014,Bruce2014}. The G13 SAM reproduces observations more closely with $50\%$ ($95\%$) of systems having B/T $< 0.5$ at $z=0$ ($z=3$). Because of this, the velocity dispersions inferred within this SAM at the highest redshifts probed can be considered discrepant with observations only at the highest redshifts. While these caveats must be taken into consideration, the values of velocity dispersion inferred are physical and generally in agreement with observations of spheroidal/passive systems \citep{Bernardi2010,Oser2012}, with $1.8 < \log \sigma_\star\ [\mathrm{km\ s}^{-1}] < 2.7$ depending on the SAM.

We must also consider whether the physical sizes of the simulated galaxy components can be used to infer the velocity dispersion. In G11 and G13, the resulting mass-size relations are shallower than the observations (see, e.g., \citealp{Lani2013,vanderwel2014}) with both masses and sizes larger (smaller) at low (high) redshift \citep{Guo2011}. For the purposes of this work however, we only care if the evolution in $M/R$ is represented correctly, and it is possible that, at least within G11 and G13, this may not be the case. We conclude that inferred velocity dispersion could provide a useful property with which to trace the evolving properties of the most massive galaxies. However, more detailed future simulations that accurately reproduce the evolution in both stellar mass and galaxy component sizes, or that report a value for velocity dispersion directly, are needed to confirm these findings.

\subsection{Comparison with previous works}
Our results are consistent with the work of \citet{Leja2013} who investigate cumulative number density selection of descendant galaxies over $0 < z < 3$ in the range $0.5 < n\ [10^{-4}\ \mathrm{Mpc}^{-3}] < 8.0 $ using the G11 SAM. Uncorrected for mergers and growth scatter, they show that for the two smallest number densities, the median stellar mass evolution is overestimated by between $0.05 - 0.15$ dex ($12-41\%$) by $z=0$. Using the mean stellar mass, we find an overestimate at $n = 5 \times 10^{-5} \mathrm{Mpc}^{-3}$ of $40\%$ in the G11 and G13 SAMs. Most recently, \citet[][H06]{Henriques2014} contrasted the mean stellar mass of progenitors derived from a constant number density selection with the values obtained from their SAM using a Planck \citep{Ade2014} cosmology. They found that mean stellar mass evolution is overestimated by a factor of $3-5$ for the progenitors of galaxies with a $z=0$ stellar mass between $10.25 < \log\ M_*\ [\mathrm{M}_\odot\ \mathrm{h}^{-2}] < 11.25$. This is a larger increase than we find for similar mass galaxies. We suggest this discrepancy is due to the ability of H06 to correctly reproduce the abundance of massive galaxies out to higher redshift. With less massive and passive galaxies, H06 must produce a larger evolution in stellar mass from high redshift to match the local stellar mass function.

\section{Summary}

We have compared the use of two popular galaxy selection methods and contrasted the use of galaxy stellar mass and inferred velocity dispersion in semi-analytical models based on the output of the Millennium Simulation over the redshift range $0 < z < 3$. We select galaxies above a constant limit of stellar mass in this redshift range, and conclude that:

\begin{itemize}
	\item Descendants can be fully recovered over the entire redshift range regardless of mass limit choice. However, progenitors of $z=0$ galaxies are lost from the selection with at least $80\%$ below the stellar mass limit at $z=1$.
    \item Average descendant ensemble stellar mass is increasingly underestimated with increasing redshift by an amount that varies between SAMs but is, on average, around $50\%$ at $z=0$. Similarly, the average stellar mass of progenitors is increasingly overestimated. At $z=0$, average progenitor stellar mass is overestimated by a factor of $\sim 5$.
\end{itemize}

Additionally, selecting galaxies at a constant cumulative number density in stellar mass, we conclude that:

\begin{itemize}
	\item Recovery of individual descendant galaxies falls exponentially with a time scale dependent on choice of number density. Just $30\%$ of the most massive galaxies (selected at $\log\ n = -4.3\ \mathrm{Mpc}^{-3}\ h^{3}$ at $z=3$) are at the same cumulative number density at the lowest redshift. For the largest number density selection, this increases to $60\%$. Recovery of progenitors is similar, but with $50 - 30\%$ recovered at the highest redshifts, depending on choice of number density.
    \item The average stellar mass of descendants is overestimated by $15\%$ ($70\%$) at the highest (smallest number densities) by $z=0$, increasing linearly from $z=3$. Furthermore, independent of number density, progenitors' average stellar mass is overestimated by $\sim50\%$ at the highest redshift.
\end{itemize}

Finally, we have investigated inferred velocity dispersion as a property with which to trace galaxies over the same redshift range. We have found that a constant number density in velocity dispersion recovers average stellar mass, stellar mass density and average SFR to within $\pm50\%$ for both descendants and progenitors. Furthermore, selecting galaxies at a constant velocity dispersion limit recovers the aforementioned properties to within $\pm80\%$ of the true values. However, these results are based on inferring velocity dispersion of galaxies which may not be strictly applicable to some SAMs and redshift ranges. Further study is required into velocity dispersion through dedicated simulations that predict these properties.

In conclusion, selecting galaxies at a constant cumulative number density is found to trace the true evolution of average stellar mass and the average SFR of the progenitors and descendants of galaxies in initial selections at $z=0$ and $z=3$. However, it does not trace the exact same galaxies but rather galaxies with very similar properties. Furthermore, it is found that selecting galaxies above a constant stellar mass with redshift returns the actual evolution within a larger factor of between two and thirty.

\section*{Acknowledgements}
We wish to thank Kenneth Duncan and Frazer Pearce for useful discussions related to this work, and Joel Leja for his rapid response in request for information. We thank the anonymous referee for their constructive comments that improved the contents of this paper. We gratefully acknowledge support from the Science and Technology Facilities Council (STFC) and the Leverhulme Trust. The Millennium Simulation databases used in this paper and the web application providing online access to them were constructed as part of the activities of the German Astrophysical Virtual Observatory (GAVO).

\bibliography{Mendeley}
\bibliographystyle{mn2e}

\label{lastpage}
\appendix

\section{Fitting Parameters}
Here we present tabulated fitting parameters for descendant and progenitor galaxy populations, as described in \S\ref{sec:fitting}. These are obtained by taking the mean of individual SAM data points and taking the variance between SAMs as the $1\sigma$ of a Gaussian distribution on each data point. Fitting is then processed on randomly sampled values $10^4$ times to calculate the most likely fitting parameters and their associated uncertainties.

\begin{table*}
\begin{minipage}{126mm}
\caption{Fitting parameters, described in \S\ref{sec:fitting}, for descendants and progenitors selected at a constant cumulative number density across the redshift range $0 < z < 3$.}
\label{tab:fitting-results-n}
	\begin{tabular}{@{}lrrrrrrr}
    
    \hline \hline
    \multirow{2}{*}{Metric} & \multicolumn{3}{c}{Descendants} & & \multicolumn{3}{c}{Progenitors} \\ \cline{2-4} \cline{6-8}
    	& $a$ & $b$ & $c$ & & $a$ & $b$ & $c$ \\
    \hline
    \multicolumn{8}{c}{$n = 1 \times 10^{-3}\ [\mathrm{Mpc}^{-3}\ h^3$]} \\
    \hline
    $f_\mathrm{rec}$ & $0.000 \pm 0.000$ & $0.480 \pm 0.030$ & $0.175 \pm 0.018$ & & $0.000 \pm 0.000$ & $1.212 \pm 0.055$ & $-0.260 \pm 0.027$ \\
    $f_\mathrm{contam}$\footnote{Descendants use Equation \ref{eqn:recfracformfit} and progenitors use Equation \ref{eqn:contamprogformfit} for fitting.} & $0.511 \pm 0.019$ & $-0.010 \pm 0.001$ & $1.000 \pm 0.000$ & & $-0.708 \pm 0.070$ & $-0.235 \pm 0.056$ & $0.563 \pm 0.064$ \\
    $\kappa_{m^*}$ & $0.191 \pm 0.066$ & $-0.044 \pm 0.019$ & - & & $-0.131 \pm 0.030$ & $0.125 \pm 0.017$ & - \\
    $\kappa_{\rho^*}$ & $0.592 \pm 0.050$ & $-0.156 \pm 0.015$ & - & & $-0.132 \pm 0.031$ & $0.126 \pm 0.018$ & - \\
    
    \hline
    \multicolumn{8}{c}{$n = 5 \times 10^{-4}\ [\mathrm{Mpc}^{-3}\ h^3$]} \\
    \hline
    $f_\mathrm{rec}$ & $0.000 \pm 0.000$ & $0.372 \pm 0.044$ & $0.240 \pm 0.032$ & & $0.000 \pm 0.000$ & $1.234 \pm 0.057$ & $-0.302 \pm 0.029$ \\
    $f_\mathrm{contam}$ & $0.568 \pm 0.032$ & $-0.011 \pm 0.001$ & $1.000 \pm 0.000$ & & $-0.670 \pm 0.035$ & $-0.242 \pm 0.017$ & $0.580 \pm 0.009$ \\
	$\kappa_{m^*}$ & $0.294 \pm 0.070$ & $-0.071 \pm 0.020$ & - & & $-0.134 \pm 0.058$ & $0.134 \pm 0.036$ & - \\
    $\kappa_{\rho^*}$ & $0.596 \pm 0.061$ & $-0.156 \pm 0.018$ & - & & $-0.134 \pm 0.058$ & $0.134 \pm 0.036$ & - \\
    
    \hline
    \multicolumn{8}{c}{$n = 1 \times 10^{-4}\ [\mathrm{Mpc}^{-3}\ h^3$]} \\
    \hline
    $f_\mathrm{rec}$ & $0.000 \pm 0.000$ & $0.180 \pm 0.047$ & $0.430 \pm 0.071$ & & $0.000 \pm 0.000$ & $1.516 \pm 0.155$ & $-0.519 \pm 0.079$ \\
	$f_\mathrm{contam}$ & $0.725 \pm 0.050$ & $-0.014 \pm 0.001$ & $1.000 \pm 0.000$ & & $-0.623 \pm 0.036$ & $-0.216 \pm 0.019$ & $0.582 \pm 0.005$ \\
    $\kappa_{m^*}$ & $0.691 \pm 0.099$ & $-0.151 \pm 0.029$ & - & & $-0.107 \pm 0.072$ & $0.144 \pm 0.045$ & - \\
    $\kappa_{\rho^*}$ & $0.742 \pm 0.062$ & $-0.194 \pm 0.018$ & - & & $-0.107 \pm 0.073$ & $0.145 \pm 0.046$ & - \\
    
    \hline
    \multicolumn{8}{c}{$n = 5 \times 10^{-5}\ [\mathrm{Mpc}^{-3}\ h^3$]} \\
    \hline
    $f_\mathrm{rec}$ & $0.000 \pm 0.000$ & $0.128 \pm 0.044$ & $0.522 \pm 0.096$ & & $0.000 \pm 0.000$ & $1.723 \pm 0.321$ & $-0.626 \pm 0.144$ \\
    $f_\mathrm{contam}$ & $0.787 \pm 0.054$ & $-0.014 \pm 0.001$ & $1.000 \pm 0.000$ & & $-0.621 \pm 0.058$ & $-0.204 \pm 0.026$ & $0.582 \pm 0.009$ \\
    $\kappa_{m^*}$ & $0.747 \pm 0.136$ & $-0.198 \pm 0.041$ & - & & $-0.083 \pm 0.062$ & $0.139 \pm 0.036$ & - \\
    $\kappa_{\rho^*}$ & $0.868 \pm 0.093$ & $-0.228 \pm 0.028$ & - & & $-0.085 \pm 0.064$ & $0.141 \pm 0.037$ & - \\
    
    \hline \hline

	\end{tabular}
\end{minipage}
\end{table*}

\end{document}